\documentclass[aps,twocolumn,prl,showpacs,floatfix,superscriptaddress]{revtex4}

\usepackage{amsmath,fixmath}
\usepackage{graphicx}
\usepackage{dcolumn}
\usepackage{amssymb}
\usepackage{color}
\usepackage{natbib}
\usepackage[T1]{fontenc}
\usepackage{textcomp}
\usepackage{txfonts}
\usepackage{hyperref}

\begin{document}

\title{Ultrametricity and long-range correlations in the Edwards-Anderson spin glass}

\author{A.~Maiorano} \affiliation{Dipartimento di Fisica, La Sapienza Universit\`a di Roma,
     00185 Roma,  Italy.}\affiliation{Instituto de Biocomputaci\'on y F\'{\i}sica de Sistemas
  Complejos (BIFI), 50009 Zaragoza, Spain.}
\author{G.~Parisi}\affiliation{Dipartimento di Fisica, IPCF-CNR, UOS
Roma Kerberos and INFN, La Sapienza Universit\`a di Roma, 00185 Roma,  Italy.}
\author{D.~Yllanes}\affiliation{Dipartimento di Fisica, La Sapienza Universit\`a di Roma,
     00185 Roma,  Italy.}  \affiliation{Instituto de
  Biocomputaci\'on y F\'{\i}sica de Sistemas Complejos (BIFI), 50009 Zaragoza,
  Spain.}

\date{\today}

\begin{abstract}
In recent times, the theoretical study of the three-dimensional
Edwards-Anderson model has produced several rigorous results on the nature of
the spin-glass phase. In particular, it has been shown that, as soon as the
overlap distribution is non-trivial, ultrametricity holds. However, these
theorems are valid only in the thermodynamical limit and are therefore of
uncertain applicability for (perennially off-equilibrium) experimental spin
glasses. In addition, their basic assumption of non-triviality is still hotly
debated. This paper intends to show that the predictions stemming from ultrametricity are
already well satisfied for the lattice sizes where numerical simulations are
possible (i.e., up to $V = 32^3$ spins) and are, therefore, relevant at
experimental scales. To this end we introduce a three-replica correlation
function, which evinces the ultrametric properties of the system and is shown
to scale in the same way as the overlap correlation function.
\end{abstract}
\maketitle 


During the last decade, the understanding of the properties of the
low-temperature phase of model spin
glasses~\cite{mezard:87,fischer:93,young:97} has made significant progress,
thanks both to theoretical advances and numerical simulations.  The Mean Field
solution~\cite{parisi:79,parisi:79b,parisi:80} is known since the eighties, but
its relatively recent rigorous proof has been lacking for more than twenty
years~\cite{talagrand:06}.  The debate remains~\cite{marinari:00,moore:05}
whether the peculiar features of the Mean Field solution are present in
realistic, finite-dimensional model spin glasses (the \emph{Replica Symmetry
Breaking} scenario, RSB) or whether a completely different picture, the
\emph{droplet model}~\cite{mcmillan:84,fisher:86,fisher:88b,bray:87} describes
the spin-glass phase.  Indeed, the central issue of whether the spin-glass order
parameter has a non-trivial distribution is still very much the subject of
active discussion (see, e.g., \cite{contucci:07b,janus:10,yucesoy:12,billoire:13,middleton:13}
for recent examples). Thus, the detailed investigation into the properties of
the spin-glass phase remains an active field.

In this paper, we build on recent advances in the study of the structure
of correlations in the spin-glass phase~\cite{contucci:09,janus:10,janus:10b}
in order to test one of the most conspicuous features of the RSB picture:
the ultrametric structure of the low-temperature phase. We shall define 
a (would-be) ultrametric correlation function and show that it scales
just as the standard spin autocorrelation, validating the prediction
of the RSB theory. To this
end we shall take advantage of the unprecedented statistics afforded to
us by the use of the Janus computer~\cite{janus:08,janus:09,janus:12b}.

In what follows we consider the Edwards-Anderson model, a long-studied
paradigm for realistic spin glasses:
\begin{equation}
H=-\sum_{\langle i,j \rangle} J_{ij}S_iS_j
\label{eq:EA}
\end{equation}
where $S_i$ are Ising spins and $J_{ij}$ are i.i.d. random quenched couplings
between nearest-neighbor sites $i,j$ on a finite-dimensional cubic lattice.  We
define as usual the total overlap of an equilibrium configuration at a given
temperature of model~(\ref{eq:EA}) as the microscopic average of local
(single-site) overlaps $q(i)=S_i^aS_i^b$, $q=\bigl[\sum_iq(i)\bigr]/V$ where
$i$ is a cubic lattice site label, $V=L^3$ is the system volume and $a,b$ are
labels for two independent replicas of the system.  This model undergoes a
second-order phase transition~\cite{gunnarsson:91,ballesteros:00,palassini:99}
at temperature $T_\text{c}=1.1019(29)$~\cite{janus:13b}.  The RSB and
droplet pictures provide very different descriptions of the $T<T_\text{c}$
spin-glass phase.


In the droplet model, the low-temperature phase is governed by a single pair
of states (related by a global spin inversion) and excitations are produced
by coherently flipping compact regions. If $\ell$ is the typical
size of such \emph{droplets}, the energy of the excitations grows as a power of
$\ell$, making system-wide excitations unaccessible in the thermodynamic limit.
All peculiar dynamical and equilibrium features of the spin-glass phase 
come from the complex interaction of droplet excitations.
In the off-equilibrium regime, the spin-glass order builds in a
super-universal coarsening dynamics~\cite{fisher:88b}.
The order parameter of the spin-glass transition is the overlap, whose value
is well defined below the transition
temperature so the  probability distribution $P(q)$ in the thermodynamic
limit is a pair of delta functions:  $P(q)=\delta(q^2-q_{\text{EA}}^2)$. 
The introduction of any external driving field completely destroys the spin-glass
phase and the system is paramagnetic at all $T>0$.

In the Replica Symmetry Breaking scenario, infinitely many states contribute to
the thermodynamics; excitations cost a finite amount of energy and fill all the
available space~\cite{marinari:00,billoire:12}.  The probability distribution
of the overlap at a given non-zero temperature in the spin-glass phase has a
delta function at $q=\pm q_{\text{EA}}$ as well as a finite weight down to
$q=0$.  The probability distribution of the overlaps is strongly constrained by
the requirement of stochastic
stability~\cite{iniguez:96,aizenman:98,ghirlanda:98,parisi:98}.  The latter has
been shown to be a quite general property: in the case of the the
Edwards-Anderson model it has been both proved~\cite{contucci:03} and observed
numerically~\cite{janus:11}. As a consequence of a very general theorem of
Panchenko~\cite{panchenko:13}, the many states are hierarchically organized and
the phase space is ultrametric: if we take the overlap as a measure of distance
between states, and we pick three equilibrium configurations at random, they
always form an isosceles triangle. Their probability distribution, including
the  fraction of equilateral triangles is fixed by stochastic stability.


The differences in the droplet and RSB pcitures reflect
on their predictions for long-range correlations.
In what follows we are interested in space correlation functions of local
overlaps.
The usual non-connected overlap-overlap correlation function is
\begin{equation}
C_q(r)=\overline {\langle q(i)q(i+r) \rangle}\mbox{ ,}
\label{eq:C4}
\end{equation}
where $\overline{(\dots)}$ denotes the average over all disorder samples 
and $\langle (\dots) \rangle$ the thermal average for a single sample.
The correlation function at a fixed value of $q$ 
\begin{equation}
C_q(r|q=Q)=\frac{\overline{\langle q(i)q(i+r)\delta(q-Q) \rangle}}
{\overline{\langle \delta(q-Q) \rangle}}
\label{eq:C4Q}
\end{equation}
decays with a power law at long distance so that
\begin{equation}
C_q(r|q=Q) \sim Q^2 + A(Q)r^{-\theta(Q)},\qquad Q \leq q_\text{EA}.
\label{eq:C4CTHETA}
\end{equation}
with non-negative $\theta$ at all values of $Q$ up to $q_{\text{EA}}=\overline{\langle S_i
  \rangle^2}$ (and $\theta<3$)~\cite{dedominicis:98,temesvari:02,contucci:09,dedominicis:06}.

On the other hand, for $Q>q_\text{EA}$  the system is in a very forced state
and the correlations decrease exponentially, characterized by a correlation
length $\xi_q$. In the large-$L$ limit, the crossover between these two regimes
becomes a phase transition when $Q\to q_\text{EA}$ from above: $\xi_Q \propto
(Q^2-Q_\text{EA})^{-\hat \nu}$. Finally, the exponents $\hat \nu$ and
$\theta(Q=q_\text{EA})$ are related by a hyperscaling law:
$\theta(q_\text{EA})=2/\hat \nu$~\cite{janus:10b}.

The droplet and RSB pictures agree on the above description, but differ on the
shape of $\theta(Q)$ for $q<q_\text{EA}$.  In the mean field
theory~\cite{dedominicis:98} we expect $\theta(Q)$ to be a non-trivial function of $Q$.
Above the upper critical dimension $D>D_\text{u}=6$, a zero-loop computation starting
from the Mean Field approximation predicts three distinct values of the
correlation exponent in the sectors $Q=q_{\text{EA}}$ ($\theta=D-2$), $Q=0$
($\theta=D-4$) and $0<Q<q_\text{EA}$ ($\theta=D-3$). Below $D_\text{u}$, these
expectation should renormalize (in fact, the given exponents become
inconsistent with the clustering property below $D=4$: for any choice of $Q$ we
must have a correlation function decaying to a well defined value). Therefore,
in $D=3$ connected correlation functions should decay as in
Eq.~\eqref{eq:C4CTHETA} but little can be said a priori on the shape of
$\theta(Q)$ for $Q<q_{\text{EA}}$, other than it should be strictly positive.
At the critical temperature the exponent is discontinuous and
$\theta_{T_\text{c}}(0)=1+\eta$, where $\eta$ is the anomalous
dimension~\cite{dedominicis:98,dedominicis:93,parisi:97}.

\begin{figure}
\includegraphics[height=\columnwidth,angle=270]{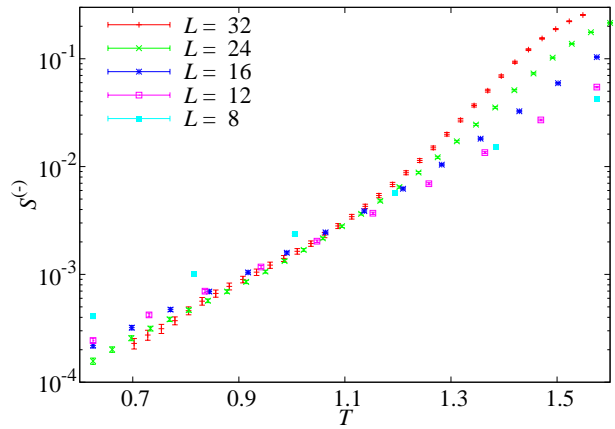}
\caption{Weight $S^{(-)}$ of the triplets where the smallest overlap is negative, 
as defined in Eq.~\eqref{eq:Smenos}. As we can see, for low temperatures the weight
of these triplets is extremely small and decreases with increasing lattice size.
\label{FIG_1}}
\end{figure}

In the droplet picture we expect a completely different scenario.
There is a unique state (apart from time-reversal symmetry)
in the thermodynamic limit with $q=q_{\text{EA}}$ at any $T<T_\text{c}$.
The space correlation function in the small-$Q$ sectors behaves as
$q_{\text{EA}}^2g(r/L)$ with $g$ a scaling function of order $\sim 1$ in the
intermediate-distance region $1\ll r \ll L$ where $L$ is the typical linear
sizes of coexistent droplets of the two symmetric phases. Therefore $\theta(Q<q_\text{EA})=0$
in the droplet picture.
For $Q=q_{\text{EA}}$, the connected correlation function decays to zero and the
power is given by the \emph{stiffness exponent} $\theta(q_\text{EA})=y$~\cite{bray:87}, whose value
has been computed to be $y=0.24(1)$ in $D=3$ (from $T=0$ studies \cite{boettcher:04}).

Recently, the use of the Janus computer~\cite{janus:12b} has permitted
a detailed numerical study of $\theta(Q)$ in $D=3$ both from
equilibrium and off-equilibrium simulations (see Sections 10.8 and 11.3
in~\cite{yllanes:11} for a complete and self-contained overview).  In
particular, for $T<T_\text{c}$ the replicon exponent has been measured to be
$\theta(0)=0.38(2)$ \cite{janus:08b,janus:09b,janus:10}, while $\hat
\nu=0.39(5)$~\cite{janus:10b}. A direct computation at $Q=q_\text{EA}$ is more
delicate, due to the stronger finite-size effects, the best value being
$\theta(q_\text{EA})= 0.61(8)$~\cite{yllanes:11}.  These numbers 
for $\theta(0)$ and $\theta(q_\text{EA})$ are in
disagreement with the droplet theory. 

In addition, $\theta(Q)$ has been seen to be constant for a finite $Q$
interval~\cite{yllanes:11}. Notice that, if indeed
$\theta(Q<q_\text{EA})=\theta(0)$, a simple Landau-like argument then implies
that $\theta(0)+1/\hat\nu = \theta(q_\text{EA})$ which, coupled with the scaling
law quoted above for $\hat\nu$, gives $\theta(q_\text{EA})=2\theta(0)=2/\hat\nu$.
The numerical results are compatible with these relations, even if our
precision is still limited.
\begin{figure}[t]
\includegraphics[width=\linewidth]{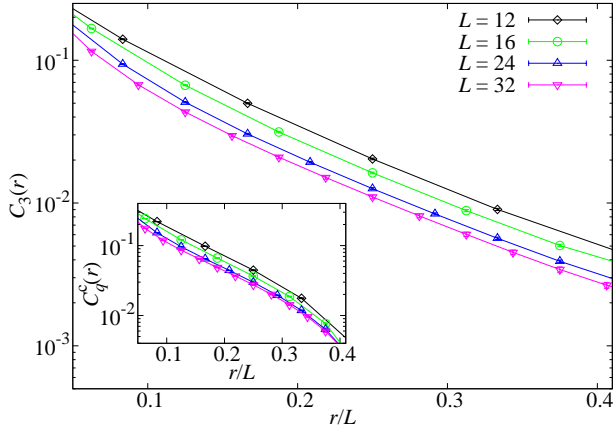}
\caption{The three-replica correlation function $C_3(r)$ for several
  system sizes at $T=0.703$. Lines are only a guide to the eye.
The inset shows the corresponding plot for the connected 
overlap correlation function $C_q^\text{c}$, which has a similar behavior 
(see, e.g., \cite{janus:10}).}
\label{FIG_2}
\end{figure}
\begin{figure}[t!]
\includegraphics[width=.9\columnwidth,angle=0]{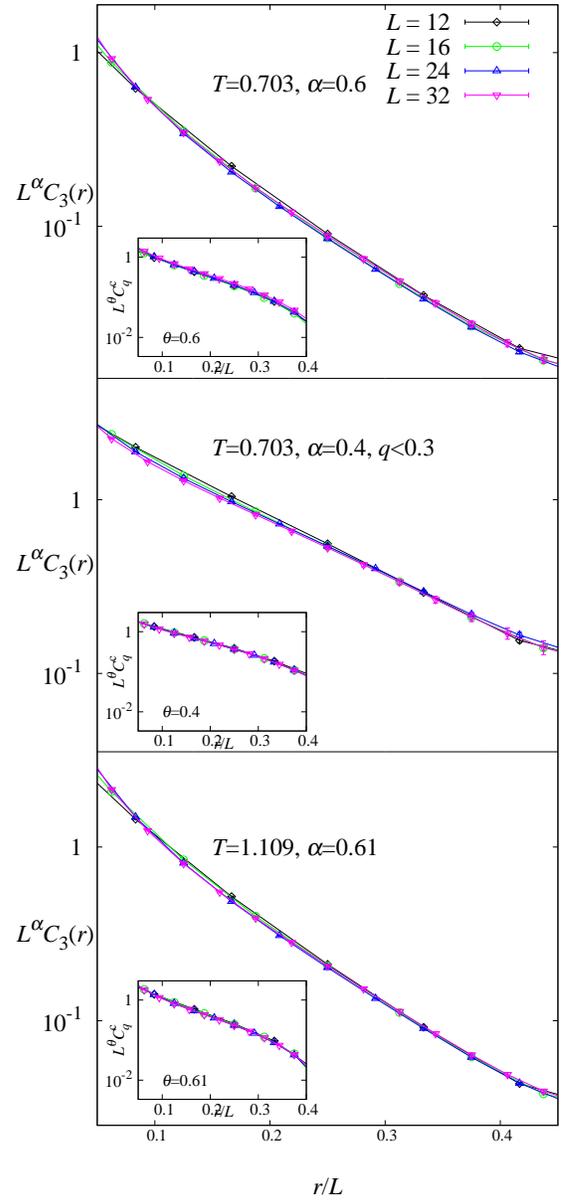}
\caption{Scaling of the three-replica corelation function $C_3$ of~\eqref{eq:C3} at $T=0.703$.
In the top panel, we show  the $C_3$ computed with all 
the triplets of configurations, which scales with an exponent $\alpha\approx0.6$.
These data are just the same as those in Figure~\ref{FIG_1}, rescaled with $\alpha$.
In the middle panel, we show that if we recompute $C_3$ only for 
triplets where all the $q^{ij}<0.3$, then $\alpha \approx 0.4$. 
Finally, at $T_\text{c}$, $\alpha$ is compatible with $1+\eta=0.6100(36)$~\cite{janus:13b}.
This behavior is compatible with that of the exponent $\theta(Q)$ that 
controls the scaling of the connected spin overlap function $C_q^\text{c}$, as shown in the 
insets. In particular, notice that the most recent 
computation gives $\theta(Q<0.3) = 0.38(2)$,
with $\theta(Q=q_\text{EA})=2 \theta(Q=0)$~\cite{janus:10b,yllanes:11}.
\label{FIG_3}}
\end{figure}
\begin{figure}[t]
\includegraphics[height=\columnwidth,angle=270]{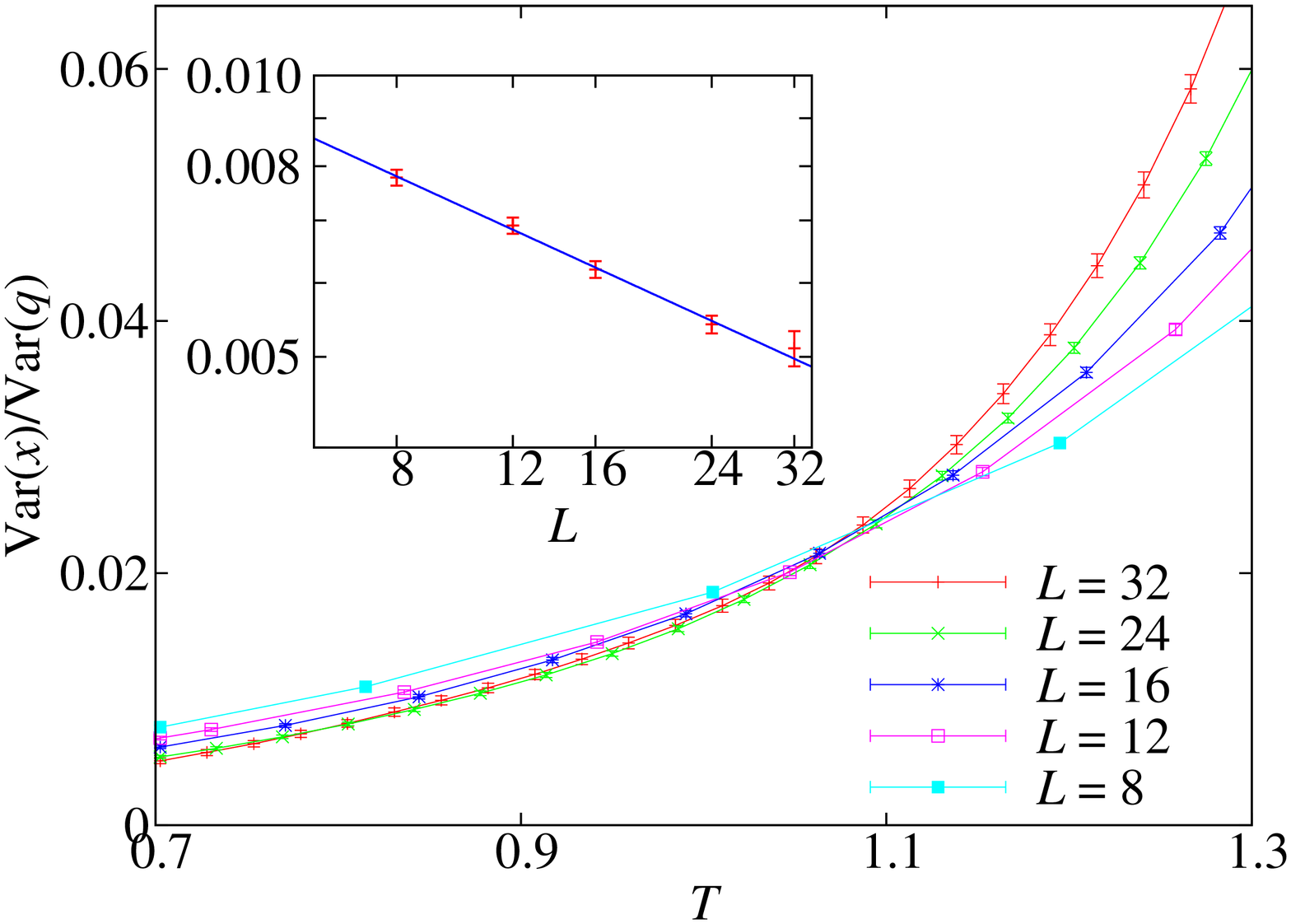}
\caption{Plot of the variance of the difference between
the two smallest overlaps $x=q^{ac}-q^{bc}$, rescaled 
by $\text{Var(q)}$. Below $T_\text{c}$ the distribution 
of $x$ goes to zero with $L$ faster than can be explained just 
from the narrowing of the peaks in $p(q)$. \textit{Inset:} 
$\text{Var}(x)/\text{Var}(q)$ as a function of $L$ at $T=0.703$,
with a fit to a power law with exponent $c=0.33(4)$. 
\label{FIG_4}}
\end{figure}

In short, thus far the numerical study of connected correlation in $D=3$ 
seems to agree with the RSB phase. Proceeding from this observation, we aim
to construct an ultrametric correlation function and study its behavior 
in terms of the scaling of $C_q(r)$. To this end, we shall analyze 
the EA model in $D=3$ defined in~\eqref{eq:EA} for lattices 
up to $L=32$ at a temperature down to $T=0.703\approx 0.64T_\text{c}$
(see~\cite{janus:10} for full details of our simulations).

Our first step is introducing some definitions.  Consider three independent
equilibrium configurations. If we use $a,b,c$ as replica labels, we can flip
configuration until the two largest overlaps are non-negative, say $q^{ab} \geq
q^{ac} \geq 0$.  This can always be done, thanks to the time-reversal symmetry
of the model.  Now, according to ultrametricity $q^{ac}=q^{bc}$, so  in the
thermodynamical limit the third overlap should also be positive and equal to
the second largest one. This equality is obviously not verified for finite
systems, but we expect that the probability that $q^{bc}<0$ be small
for large volume sizes~\cite{contucci:07b}.

This last point is crucial: if the weight of the triangles with a negative
side were significative, the rest of our analysis would not rest on 
a solid foundation (the system would either not be ultrametric or 
we would be too far from the asymptotic regime for our results to have any value).
In order to test this issue we define the following quantity:
\begin{equation}\label{eq:Smenos}
S^{(-)} = \frac{\int_{-1}^0 \text{d} q^{bc} \ (q^{bc})^2 p(q^{bc})}{\int_{-1}^1 \text{d} q^{bc} \ (q^{bc})^2 p(q^{bc})}.
\end{equation}
As we can see in Figure~\ref{FIG_1}, the value of $S^{(-)}$ is indeed very small 
below the critical temperature and, furthermore, it decreases with increasing
lattice size. Notice that $S^{(-)}$ decreases very quickly (exponentially) when we decrease
the temperature. Therefore, in what follows we shall work at the lowest temperature
for which we have data up to $L=32$, $T=0.703$.

As we have said, ultrametricity requires that triangles must be isosceles and
$q^{bc}=q^{ac}$. It is thefore interesting to consider the difference
\begin{equation}\label{eq:x} x = q^{ac}-q^{bc} = \frac{1}{V} \sum_i q^{ac}_i - q^{bc}_i = \frac1V \sum_i x_i. \end{equation}
In particular, we define 
\begin{eqnarray}
Q_u^2 & = & \overline{\langle(q^{bc}-q^{ac})^2\rangle} = \overline{\langle x^2\rangle} \nonumber \\
 & = & \frac{1}{V^2}\sum_i\sum_k   
\overline{\langle x_i x_k\rangle}\mbox{ ,}
\label{eq:Qu2}
\end{eqnarray}
which should vanish in the thermodynamical limit.
We then define a three-replica correlation function
from the autocorrelation of $x$ (analogous to $C_q$):
\begin{eqnarray}
C_3(r) & = &  \frac{1}{V}\sum_i \overline{\langle x_i x_{i+r}\rangle},
\label{eq:C3}
\end{eqnarray}
which verifies $Q_u^2 = \frac{1}{V}\sum_rC_3(r)$.

A rapidly vanishing $C_3(r)$ at large distance would then be a signature of
ultrametricity. We are tempted to conjecture that the behavior is not
dissimilar from the one of the connected overlap-overlap correlation function,
and that at long distance $C_3(r)/C_q^\text{c}(r) \sim O(1)$, 
where $C_q^\text{c}(r)=C_q(r)-\overline{\langle q^2\rangle}$ is the connected
version of~\eqref{eq:C4}.  In the droplet
picture, since no states with $q\neq q_\text{EA}$ survive in the thermodynamic
limit, all triangles all equilateral and $Q_u^2$ is trivially null.  


We show data for $C_q^\text{c}$ and $C_3$
in Figure~\ref{FIG_2} for various system sizes at
temperature $T=0.703$. Both functions decay to zero at large distances.
We can now look for an algebraic decay of the form
\begin{equation}
C_3(r)=\frac{1}{r^\alpha}f(r/L).
\end{equation}
We have attempted this in Figure~\ref{FIG_3}. In 
the upper panel we show $C_3(r)$ computed with all the triplets 
of configurations. We are able to obtain a reasonably good 
collapse of the data for the largest system sizes with 
$\alpha \approx 0.6$.  However, should $C_3$  really 
scale as $C_q$, we expect that number to be only 
an effective exponent, combining the effect of the 
different $q$ sectors.

In principle, we would like to study the dependence of $\alpha$ on $Q$ 
and, in particular, whether $\alpha(Q)=\theta(Q)$.
Unfortunately, since $C_3$ is a three-replica function  we cannot 
write it as a function of a single overlap, as in~\eqref{eq:C4Q}. 
However, recall the numerical 
observation that $\theta(Q<\mathcal C)=\theta(0)$, 
where $C$ is a finite cutoff value (expected to be $\mathcal C=q_\text{EA}$
in the thermodynamical limit, but $\mathcal C\approx 0.3$ for our system
sizes~\cite{yllanes:11}). Assuming that $\alpha$  has a similar
behavior we have recomputed $C_3$ considering
only the triplets where all the $q^{ij}$ are smaller than $\mathcal C=0.3$. 
We see in the middle panel of Figure~\ref{FIG_2} that 
now the value of $\alpha$ that produces the best collapse 
is $\alpha\approx 0.4$, compatible with the value $\theta(0)=0.38(2)$
found for $C_q$.

In the lower panel of Figure~\ref{FIG_3} we also show collapsed data at a
$T=T_\text{c}$.  In this case we do not need to impose any cutoff 
and the collapse for $L>12$ is compatible with the 
ansatz $\alpha_{T_\text{c}}= \theta_{T_\text{c}} = 1 + \eta$, with
$\eta=-0.3900(36)$ from~\cite{janus:13b}.


We can get more information on the distribution of triplets of configurations
with ordered overlaps from a study of the probability 
distribution of $x$, which should approach a delta function as the system
size increases.  In order to test this hypothesis, we can study the variance of
$x$.  We have represented this quantity in Figure~\ref{FIG_4}, normalized by
$\text{Var}(q)=\overline{\langle q^2\rangle}$ (this is to absorb the effect of the narrowing peaks in the
$p(q)$ as the system size grows). Considered as a function of $L$,
$\text{Var}(x)/\text{Var}(q)$ has a clearly different behavior at low and
high temperature. Below $T_\text{c}$ we can see that $\text{Var}(x)$ decreases
with $L$ at a rate that cannot be explained simply by a narrowing of the $q$
distribution.  Indeed, if we consider $T=0.703$ and fit
$\text{Var}(x)/\text{Var}(q) = A L^{-c}$, we obtain a value of $c=0.33(4)$,
with $\chi^2/\text{d.o.f.}=0.84/3$. Notice that in the thermodynamical limit
$\text{Var}(q)$ is finite, while, according to the previous study,
we should expect $\text{Var}(x)$ to decay algebraically with an exponent
$\alpha = \theta(0)=0.38(2)$, which is very close to the value of $c$ 
from the fit (there are probably some preasymptotic effects due to the narrowing
of the $q$ distribution).  This is a clear
quantitative sign that all the overlap triangles are isosceles (or
equilateral) in the thermodynamical limit, but not in a trivial way.


In conclusion, we have presented an analysis of statistics taken from triplets of
independent configurations and found clear signatures of ultrametricity.
We introduce a three-replica ultrametric correlation function that decays algebraically
with distance with an exponent compatible with the predictions of the RSB theory.
In the thermodynamic limit it is always possible to flip configurations to
have contributions only from non-frustrated triplets.
The variance of the difference between minimum and mid-value overlap in these triplets is
vanishing in the thermodynamic limit.

\begin{acknowledgments}
We thank the Janus Collaboration for allowing us to analyze their
thermalized configurations.
This research has been supported by the European Research
Council under the European Union's Seventh Framework Programme (FP7/2007-2013,
ERC grant agreement no. 247328). DY acknowledges support from MINECO, Spain
(grant agreement no. FIS2012-35719-C02).
\end{acknowledgments}


\begin{thebibliography}{44}%
\makeatletter
\providecommand \@ifxundefined [1]{%
 \@ifx{#1\undefined}
}%
\providecommand \@ifnum [1]{%
 \ifnum #1\expandafter \@firstoftwo
 \else \expandafter \@secondoftwo
 \fi
}%
\providecommand \@ifx [1]{%
 \ifx #1\expandafter \@firstoftwo
 \else \expandafter \@secondoftwo
 \fi
}%
\providecommand \natexlab [1]{#1}%
\providecommand \enquote  [1]{``#1''}%
\providecommand \bibnamefont  [1]{#1}%
\providecommand \bibfnamefont [1]{#1}%
\providecommand \citenamefont [1]{#1}%
\providecommand \href@noop [0]{\@secondoftwo}%
\providecommand \href [0]{\begingroup \@sanitize@url \@href}%
\providecommand \@href[1]{\@@startlink{#1}\@@href}%
\providecommand \@@href[1]{\endgroup#1\@@endlink}%
\providecommand \@sanitize@url [0]{\catcode `\\12\catcode `\$12\catcode
  `\&12\catcode `\#12\catcode `\^12\catcode `\_12\catcode `\%12\relax}%
\providecommand \@@startlink[1]{}%
\providecommand \@@endlink[0]{}%
\providecommand \url  [0]{\begingroup\@sanitize@url \@url }%
\providecommand \@url [1]{\endgroup\@href {#1}{\urlprefix }}%
\providecommand \urlprefix  [0]{URL }%
\providecommand \Eprint [0]{\href }%
\providecommand \doibase [0]{http://dx.doi.org/}%
\providecommand \selectlanguage [0]{\@gobble}%
\providecommand \bibinfo  [0]{\@secondoftwo}%
\providecommand \bibfield  [0]{\@secondoftwo}%
\providecommand \translation [1]{[#1]}%
\providecommand \BibitemOpen [0]{}%
\providecommand \bibitemStop [0]{}%
\providecommand \bibitemNoStop [0]{.\EOS\space}%
\providecommand \EOS [0]{\spacefactor3000\relax}%
\providecommand \BibitemShut  [1]{\csname bibitem#1\endcsname}%
\let\auto@bib@innerbib\@empty
\bibitem [{\citenamefont {M{\'e}zard}\ \emph {et~al.}(1987)\citenamefont
  {M{\'e}zard}, \citenamefont {Parisi},\ and\ \citenamefont
  {Virasoro}}]{mezard:87}%
  \BibitemOpen
  \bibfield  {author} {\bibinfo {author} {\bibfnamefont {M.}~\bibnamefont
  {M{\'e}zard}}, \bibinfo {author} {\bibfnamefont {G.}~\bibnamefont {Parisi}},
  \ and\ \bibinfo {author} {\bibfnamefont {M.}~\bibnamefont {Virasoro}},\
  }\href@noop {} {\emph {\bibinfo {title} {Spin-Glass Theory and Beyond}}}\
  (\bibinfo  {publisher} {World Scientific},\ \bibinfo {address} {Singapore},\
  \bibinfo {year} {1987})\BibitemShut {NoStop}%
\bibitem [{\citenamefont {Fischer}\ and\ \citenamefont
  {Hertz}(1993)}]{fischer:93}%
  \BibitemOpen
  \bibfield  {author} {\bibinfo {author} {\bibfnamefont {K.~H.}\ \bibnamefont
  {Fischer}}\ and\ \bibinfo {author} {\bibfnamefont {J.~A.}\ \bibnamefont
  {Hertz}},\ }\href@noop {} {\emph {\bibinfo {title} {Spin Glasses}}}\
  (\bibinfo  {publisher} {Cambridge University Press},\ \bibinfo {year}
  {1993})\BibitemShut {NoStop}%
\bibitem [{\citenamefont {Young}(1997)}]{young:97}%
  \BibitemOpen
  \bibfield  {author} {\bibinfo {author} {\bibfnamefont {A.~P.}\ \bibnamefont
  {Young}},\ }in\ \href@noop {} {\emph {\bibinfo {booktitle} {Spin Glasses and
  Random Fields}}},\ \bibinfo {editor} {edited by\ \bibinfo {editor}
  {\bibfnamefont {A.~P.}\ \bibnamefont {Young}}}\ (\bibinfo  {publisher} {World
  Scientific},\ \bibinfo {address} {Singapore},\ \bibinfo {year}
  {1997})\BibitemShut {NoStop}%
\bibitem [{\citenamefont {Parisi}(1979{\natexlab{a}})}]{parisi:79}%
  \BibitemOpen
  \bibfield  {author} {\bibinfo {author} {\bibfnamefont {G.}~\bibnamefont
  {Parisi}},\ }\href {\doibase 10.1103/PhysRevLett.43.1754} {\bibfield
  {journal} {\bibinfo  {journal} {Phys. Rev. Lett.}\ }\textbf {\bibinfo
  {volume} {43}},\ \bibinfo {pages} {1754} (\bibinfo {year}
  {1979}{\natexlab{a}})}\BibitemShut {NoStop}%
\bibitem [{\citenamefont {Parisi}(1979{\natexlab{b}})}]{parisi:79b}%
  \BibitemOpen
  \bibfield  {author} {\bibinfo {author} {\bibfnamefont {G.}~\bibnamefont
  {Parisi}},\ }\href {\doibase 10.1016/0375-9601(79)90708-4} {\bibfield
  {journal} {\bibinfo  {journal} {Phys. Lett.}\ }\textbf {\bibinfo {volume}
  {73A}},\ \bibinfo {pages} {203} (\bibinfo {year}
  {1979}{\natexlab{b}})}\BibitemShut {NoStop}%
\bibitem [{\citenamefont {Parisi}(1980)}]{parisi:80}%
  \BibitemOpen
  \bibfield  {author} {\bibinfo {author} {\bibfnamefont {G.}~\bibnamefont
  {Parisi}},\ }\href {\doibase 10.1088/0305-4470/13/3/042} {\bibfield
  {journal} {\bibinfo  {journal} {J. Phys. A: Math. Gen.}\ }\textbf {\bibinfo
  {volume} {13}},\ \bibinfo {pages} {1101} (\bibinfo {year}
  {1980})}\BibitemShut {NoStop}%
\bibitem [{\citenamefont {Talagrand}(2006)}]{talagrand:06}%
  \BibitemOpen
  \bibfield  {author} {\bibinfo {author} {\bibfnamefont {M.}~\bibnamefont
  {Talagrand}},\ }\href@noop {} {\bibfield  {journal} {\bibinfo  {journal}
  {Ann. of Math.}\ }\textbf {\bibinfo {volume} {163}},\ \bibinfo {pages} {221}
  (\bibinfo {year} {2006})}\BibitemShut {NoStop}%
\bibitem [{\citenamefont {Marinari}\ \emph {et~al.}(2000)\citenamefont
  {Marinari}, \citenamefont {Parisi}, \citenamefont {Ricci-Tersenghi},
  \citenamefont {Ruiz-Lorenzo},\ and\ \citenamefont {Zuliani}}]{marinari:00}%
  \BibitemOpen
  \bibfield  {author} {\bibinfo {author} {\bibfnamefont {E.}~\bibnamefont
  {Marinari}}, \bibinfo {author} {\bibfnamefont {G.}~\bibnamefont {Parisi}},
  \bibinfo {author} {\bibfnamefont {F.}~\bibnamefont {Ricci-Tersenghi}},
  \bibinfo {author} {\bibfnamefont {J.~J.}\ \bibnamefont {Ruiz-Lorenzo}}, \
  and\ \bibinfo {author} {\bibfnamefont {F.}~\bibnamefont {Zuliani}},\ }\href
  {\doibase 10.1023/A:1018607809852} {\bibfield  {journal} {\bibinfo  {journal}
  {J. Stat. Phys.}\ }\textbf {\bibinfo {volume} {98}},\ \bibinfo {pages} {973}
  (\bibinfo {year} {2000})},\ \Eprint
  {http://arxiv.org/abs/arXiv:cond-mat/9906076} {arXiv:cond-mat/9906076}
  \BibitemShut {NoStop}%
\bibitem [{\citenamefont {Moore}(2005)}]{moore:05}%
  \BibitemOpen
  \bibfield  {author} {\bibinfo {author} {\bibfnamefont {M.~A.}\ \bibnamefont
  {Moore}},\ }\href {\doibase 10.1088/0305-4470/38/46/L03} {\bibfield
  {journal} {\bibinfo  {journal} {J. Phys. A.: Math. Gen.}\ }\textbf {\bibinfo
  {volume} {38}},\ \bibinfo {pages} {L783} (\bibinfo {year} {2005})},\ \Eprint
  {http://arxiv.org/abs/arXiv:cond-mat/0508087} {arXiv:cond-mat/0508087}
  \BibitemShut {NoStop}%
\bibitem [{\citenamefont {McMillan}(1984)}]{mcmillan:84}%
  \BibitemOpen
  \bibfield  {author} {\bibinfo {author} {\bibfnamefont {W.~L.}\ \bibnamefont
  {McMillan}},\ }\href {\doibase 10.1088/0022-3719/17/18/010} {\bibfield
  {journal} {\bibinfo  {journal} {J. Phys. C: Solid State Phys.}\ }\textbf
  {\bibinfo {volume} {17}},\ \bibinfo {pages} {3179} (\bibinfo {year}
  {1984})}\BibitemShut {NoStop}%
\bibitem [{\citenamefont {Fisher}\ and\ \citenamefont
  {Huse}(1986)}]{fisher:86}%
  \BibitemOpen
  \bibfield  {author} {\bibinfo {author} {\bibfnamefont {D.~S.}\ \bibnamefont
  {Fisher}}\ and\ \bibinfo {author} {\bibfnamefont {D.~A.}\ \bibnamefont
  {Huse}},\ }\href {\doibase 10.1103/PhysRevLett.56.1601} {\bibfield  {journal}
  {\bibinfo  {journal} {Phys. Rev. Lett.}\ }\textbf {\bibinfo {volume} {56}},\
  \bibinfo {pages} {1601} (\bibinfo {year} {1986})}\BibitemShut {NoStop}%
\bibitem [{\citenamefont {Fisher}\ and\ \citenamefont
  {Huse}(1988)}]{fisher:88b}%
  \BibitemOpen
  \bibfield  {author} {\bibinfo {author} {\bibfnamefont {D.~S.}\ \bibnamefont
  {Fisher}}\ and\ \bibinfo {author} {\bibfnamefont {D.~A.}\ \bibnamefont
  {Huse}},\ }\href {\doibase 10.1103/PhysRevB.38.386} {\bibfield  {journal}
  {\bibinfo  {journal} {Phys. Rev. B}\ }\textbf {\bibinfo {volume} {38}},\
  \bibinfo {pages} {386} (\bibinfo {year} {1988})}\BibitemShut {NoStop}%
\bibitem [{\citenamefont {Bray}\ and\ \citenamefont {Moore}(1987)}]{bray:87}%
  \BibitemOpen
  \bibfield  {author} {\bibinfo {author} {\bibfnamefont {A.~J.}\ \bibnamefont
  {Bray}}\ and\ \bibinfo {author} {\bibfnamefont {M.~A.}\ \bibnamefont
  {Moore}},\ }in\ \href@noop {} {\emph {\bibinfo {booktitle} {Heidelberg
  Colloquium on Glassy Dynamics}}},\ \bibinfo {series and number} {\bibinfo
  {series} {Lecture Notes in Physics}\ No.\ \bibinfo {number} {275}},\ \bibinfo
  {editor} {edited by\ \bibinfo {editor} {\bibfnamefont {J.~L.}\ \bibnamefont
  {van Hemmen}}\ and\ \bibinfo {editor} {\bibfnamefont {I.}~\bibnamefont
  {Morgenstern}}}\ (\bibinfo  {publisher} {Springer},\ \bibinfo {address}
  {Berlin},\ \bibinfo {year} {1987})\BibitemShut {NoStop}%
\bibitem [{\citenamefont {Contucci}\ \emph {et~al.}(2007)\citenamefont
  {Contucci}, \citenamefont {Giardin{\`a}}, \citenamefont {Giberti},
  \citenamefont {Parisi},\ and\ \citenamefont {Vernia}}]{contucci:07b}%
  \BibitemOpen
  \bibfield  {author} {\bibinfo {author} {\bibfnamefont {P.}~\bibnamefont
  {Contucci}}, \bibinfo {author} {\bibfnamefont {C.}~\bibnamefont
  {Giardin{\`a}}}, \bibinfo {author} {\bibfnamefont {C.}~\bibnamefont
  {Giberti}}, \bibinfo {author} {\bibfnamefont {G.}~\bibnamefont {Parisi}}, \
  and\ \bibinfo {author} {\bibfnamefont {C.}~\bibnamefont {Vernia}},\ }\href
  {\doibase 10.1103/PhysRevLett.99.057206} {\bibfield  {journal} {\bibinfo
  {journal} {Phys. Rev. Lett}\ }\textbf {\bibinfo {volume} {99}},\ \bibinfo
  {pages} {057206} (\bibinfo {year} {2007})},\ \Eprint
  {http://arxiv.org/abs/arXiv:cond-mat/0607376} {arXiv:cond-mat/0607376}
  \BibitemShut {NoStop}%
\bibitem [{\citenamefont {Alvarez~Ba{\~n}os}\ \emph
  {et~al.}(2010{\natexlab{a}})\citenamefont {Alvarez~Ba{\~n}os}, \citenamefont
  {Cruz}, \citenamefont {Fernandez}, \citenamefont {Gil-Narvion}, \citenamefont
  {Gordillo-Guerrero}, \citenamefont {Guidetti}, \citenamefont {Maiorano},
  \citenamefont {Mantovani}, \citenamefont {Marinari}, \citenamefont
  {Martin-Mayor}, \citenamefont {Monforte-Garcia}, \citenamefont
  {Mu{\~n}oz~Sudupe}, \citenamefont {Navarro}, \citenamefont {Parisi},
  \citenamefont {Perez-Gaviro}, \citenamefont {Ruiz-Lorenzo}, \citenamefont
  {Schifano}, \citenamefont {Seoane}, \citenamefont {Tarancon}, \citenamefont
  {Tripiccione},\ and\ \citenamefont {Yllanes}}]{janus:10}%
  \BibitemOpen
  \bibfield  {author} {\bibinfo {author} {\bibfnamefont {R.}~\bibnamefont
  {Alvarez~Ba{\~n}os}}, \bibinfo {author} {\bibfnamefont {A.}~\bibnamefont
  {Cruz}}, \bibinfo {author} {\bibfnamefont {L.~A.}\ \bibnamefont {Fernandez}},
  \bibinfo {author} {\bibfnamefont {J.~M.}\ \bibnamefont {Gil-Narvion}},
  \bibinfo {author} {\bibfnamefont {A.}~\bibnamefont {Gordillo-Guerrero}},
  \bibinfo {author} {\bibfnamefont {M.}~\bibnamefont {Guidetti}}, \bibinfo
  {author} {\bibfnamefont {A.}~\bibnamefont {Maiorano}}, \bibinfo {author}
  {\bibfnamefont {F.}~\bibnamefont {Mantovani}}, \bibinfo {author}
  {\bibfnamefont {E.}~\bibnamefont {Marinari}}, \bibinfo {author}
  {\bibfnamefont {V.}~\bibnamefont {Martin-Mayor}}, \bibinfo {author}
  {\bibfnamefont {J.}~\bibnamefont {Monforte-Garcia}}, \bibinfo {author}
  {\bibfnamefont {A.}~\bibnamefont {Mu{\~n}oz~Sudupe}}, \bibinfo {author}
  {\bibfnamefont {D.}~\bibnamefont {Navarro}}, \bibinfo {author} {\bibfnamefont
  {G.}~\bibnamefont {Parisi}}, \bibinfo {author} {\bibfnamefont
  {S.}~\bibnamefont {Perez-Gaviro}}, \bibinfo {author} {\bibfnamefont {J.~J.}\
  \bibnamefont {Ruiz-Lorenzo}}, \bibinfo {author} {\bibfnamefont {S.~F.}\
  \bibnamefont {Schifano}}, \bibinfo {author} {\bibfnamefont {B.}~\bibnamefont
  {Seoane}}, \bibinfo {author} {\bibfnamefont {A.}~\bibnamefont {Tarancon}},
  \bibinfo {author} {\bibfnamefont {R.}~\bibnamefont {Tripiccione}}, \ and\
  \bibinfo {author} {\bibfnamefont {D.}~\bibnamefont {Yllanes}} (\bibinfo
  {collaboration} {Janus Collaboration}),\ }\href {\doibase
  10.1088/1742-5468/2010/06/P06026} {\bibfield  {journal} {\bibinfo  {journal}
  {J. Stat. Mech.}\ ,\ \bibinfo {pages} {P06026}} (\bibinfo {year}
  {2010}{\natexlab{a}})},\ \Eprint {http://arxiv.org/abs/arXiv:1003.2569}
  {arXiv:1003.2569} \BibitemShut {NoStop}%
\bibitem [{\citenamefont {Yucesoy}\ \emph {et~al.}(2012)\citenamefont
  {Yucesoy}, \citenamefont {Katzgraber},\ and\ \citenamefont
  {Machta}}]{yucesoy:12}%
  \BibitemOpen
  \bibfield  {author} {\bibinfo {author} {\bibfnamefont {B.}~\bibnamefont
  {Yucesoy}}, \bibinfo {author} {\bibfnamefont {H.~G.}\ \bibnamefont
  {Katzgraber}}, \ and\ \bibinfo {author} {\bibfnamefont {J.}~\bibnamefont
  {Machta}},\ }\href {\doibase 10.1103/PhysRevLett.109.177204} {\bibfield
  {journal} {\bibinfo  {journal} {Phys. Rev. Lett.}\ }\textbf {\bibinfo
  {volume} {109}},\ \bibinfo {pages} {177204} (\bibinfo {year} {2012})},\
  \Eprint {http://arxiv.org/abs/arXiv:1206.0783} {arXiv:1206.0783} \BibitemShut
  {NoStop}%
\bibitem [{\citenamefont {Billoire}\ \emph {et~al.}(2013)\citenamefont
  {Billoire}, \citenamefont {Fernandez}, \citenamefont {Maiorano},
  \citenamefont {Marinari}, \citenamefont {Martin-Mayor}, \citenamefont
  {Parisi}, \citenamefont {Ricci-Tersenghi}, \citenamefont {Ruiz-Lorenzo},\
  and\ \citenamefont {Yllanes}}]{billoire:13}%
  \BibitemOpen
  \bibfield  {author} {\bibinfo {author} {\bibfnamefont {A.}~\bibnamefont
  {Billoire}}, \bibinfo {author} {\bibfnamefont {L.~A.}\ \bibnamefont
  {Fernandez}}, \bibinfo {author} {\bibfnamefont {A.}~\bibnamefont {Maiorano}},
  \bibinfo {author} {\bibfnamefont {E.}~\bibnamefont {Marinari}}, \bibinfo
  {author} {\bibfnamefont {V.}~\bibnamefont {Martin-Mayor}}, \bibinfo {author}
  {\bibfnamefont {G.}~\bibnamefont {Parisi}}, \bibinfo {author} {\bibfnamefont
  {F.}~\bibnamefont {Ricci-Tersenghi}}, \bibinfo {author} {\bibfnamefont
  {J.}~\bibnamefont {Ruiz-Lorenzo}}, \ and\ \bibinfo {author} {\bibfnamefont
  {D.}~\bibnamefont {Yllanes}},\ }\href {\doibase
  10.1103/PhysRevLett.110.219701} {\bibfield  {journal} {\bibinfo  {journal}
  {Phys. Rev. Lett.}\ }\textbf {\bibinfo {volume} {110}},\ \bibinfo {pages}
  {219701} (\bibinfo {year} {2013})},\ \Eprint
  {http://arxiv.org/abs/arXiv:1211.0843} {arXiv:1211.0843} \BibitemShut
  {NoStop}%
\bibitem [{\citenamefont {Middleton}(2013)}]{middleton:13}%
  \BibitemOpen
  \bibfield  {author} {\bibinfo {author} {\bibfnamefont {A.~A.}\ \bibnamefont
  {Middleton}},\ }\href {\doibase 10.1103/PhysRevB.87.220201} {\bibfield
  {journal} {\bibinfo  {journal} {Phys. Rev. B}\ }\textbf {\bibinfo {volume}
  {87}},\ \bibinfo {pages} {220201} (\bibinfo {year} {2013})},\ \Eprint
  {http://arxiv.org/abs/arXiv:1303.2253} {arXiv:1303.2253} \BibitemShut
  {NoStop}%
\bibitem [{\citenamefont {Contucci}\ \emph {et~al.}(2009)\citenamefont
  {Contucci}, \citenamefont {Giardin{\`a}}, \citenamefont {Giberti},
  \citenamefont {Parisi},\ and\ \citenamefont {Vernia}}]{contucci:09}%
  \BibitemOpen
  \bibfield  {author} {\bibinfo {author} {\bibfnamefont {P.}~\bibnamefont
  {Contucci}}, \bibinfo {author} {\bibfnamefont {C.}~\bibnamefont
  {Giardin{\`a}}}, \bibinfo {author} {\bibfnamefont {C.}~\bibnamefont
  {Giberti}}, \bibinfo {author} {\bibfnamefont {G.}~\bibnamefont {Parisi}}, \
  and\ \bibinfo {author} {\bibfnamefont {C.}~\bibnamefont {Vernia}},\ }\href
  {\doibase 10.1103/PhysRevLett.103.017201} {\bibfield  {journal} {\bibinfo
  {journal} {Phys. Rev. Lett}\ }\textbf {\bibinfo {volume} {103}},\ \bibinfo
  {pages} {017201} (\bibinfo {year} {2009})},\ \Eprint
  {http://arxiv.org/abs/arXiv:0902.0594} {arXiv:0902.0594} \BibitemShut
  {NoStop}%
\bibitem [{\citenamefont {Alvarez~Ba{\~n}os}\ \emph
  {et~al.}(2010{\natexlab{b}})\citenamefont {Alvarez~Ba{\~n}os}, \citenamefont
  {Cruz}, \citenamefont {Fernandez}, \citenamefont {Gil-Narvion}, \citenamefont
  {Gordillo-Guerrero}, \citenamefont {Guidetti}, \citenamefont {Maiorano},
  \citenamefont {Mantovani}, \citenamefont {Marinari}, \citenamefont
  {Martin-Mayor}, \citenamefont {Monforte-Garcia}, \citenamefont
  {Mu{\~n}oz~Sudupe}, \citenamefont {Navarro}, \citenamefont {Parisi},
  \citenamefont {Perez-Gaviro}, \citenamefont {Ruiz-Lorenzo}, \citenamefont
  {Schifano}, \citenamefont {Seoane}, \citenamefont {Tarancon}, \citenamefont
  {Tripiccione},\ and\ \citenamefont {Yllanes}}]{janus:10b}%
  \BibitemOpen
  \bibfield  {author} {\bibinfo {author} {\bibfnamefont {R.}~\bibnamefont
  {Alvarez~Ba{\~n}os}}, \bibinfo {author} {\bibfnamefont {A.}~\bibnamefont
  {Cruz}}, \bibinfo {author} {\bibfnamefont {L.~A.}\ \bibnamefont {Fernandez}},
  \bibinfo {author} {\bibfnamefont {J.~M.}\ \bibnamefont {Gil-Narvion}},
  \bibinfo {author} {\bibfnamefont {A.}~\bibnamefont {Gordillo-Guerrero}},
  \bibinfo {author} {\bibfnamefont {M.}~\bibnamefont {Guidetti}}, \bibinfo
  {author} {\bibfnamefont {A.}~\bibnamefont {Maiorano}}, \bibinfo {author}
  {\bibfnamefont {F.}~\bibnamefont {Mantovani}}, \bibinfo {author}
  {\bibfnamefont {E.}~\bibnamefont {Marinari}}, \bibinfo {author}
  {\bibfnamefont {V.}~\bibnamefont {Martin-Mayor}}, \bibinfo {author}
  {\bibfnamefont {J.}~\bibnamefont {Monforte-Garcia}}, \bibinfo {author}
  {\bibfnamefont {A.}~\bibnamefont {Mu{\~n}oz~Sudupe}}, \bibinfo {author}
  {\bibfnamefont {D.}~\bibnamefont {Navarro}}, \bibinfo {author} {\bibfnamefont
  {G.}~\bibnamefont {Parisi}}, \bibinfo {author} {\bibfnamefont
  {S.}~\bibnamefont {Perez-Gaviro}}, \bibinfo {author} {\bibfnamefont {J.~J.}\
  \bibnamefont {Ruiz-Lorenzo}}, \bibinfo {author} {\bibfnamefont {S.~F.}\
  \bibnamefont {Schifano}}, \bibinfo {author} {\bibfnamefont {B.}~\bibnamefont
  {Seoane}}, \bibinfo {author} {\bibfnamefont {A.}~\bibnamefont {Tarancon}},
  \bibinfo {author} {\bibfnamefont {R.}~\bibnamefont {Tripiccione}}, \ and\
  \bibinfo {author} {\bibfnamefont {D.}~\bibnamefont {Yllanes}} (\bibinfo
  {collaboration} {Janus Collaboration}),\ }\href {\doibase
  10.1103/PhysRevLett.105.177202} {\bibfield  {journal} {\bibinfo  {journal}
  {Phys. Rev. Lett.}\ }\textbf {\bibinfo {volume} {105}},\ \bibinfo {pages}
  {177202} (\bibinfo {year} {2010}{\natexlab{b}})},\ \Eprint
  {http://arxiv.org/abs/arXiv:1003.2943} {arXiv:1003.2943} \BibitemShut
  {NoStop}%
\bibitem [{\citenamefont {Belletti}\ \emph
  {et~al.}(2008{\natexlab{a}})\citenamefont {Belletti}, \citenamefont
  {Cotallo}, \citenamefont {Cruz}, \citenamefont {Fernandez}, \citenamefont
  {Gordillo}, \citenamefont {Maiorano}, \citenamefont {Mantovani},
  \citenamefont {Marinari}, \citenamefont {Martin-Mayor}, \citenamefont
  {Mu{\~n}oz~Sudupe}, \citenamefont {Navarro}, \citenamefont {Perez-Gaviro},
  \citenamefont {Ruiz-Lorenzo}, \citenamefont {Schifano}, \citenamefont
  {Sciretti}, \citenamefont {Tarancon}, \citenamefont {Tripiccione},\ and\
  \citenamefont {Velasco}}]{janus:08}%
  \BibitemOpen
  \bibfield  {author} {\bibinfo {author} {\bibfnamefont {F.}~\bibnamefont
  {Belletti}}, \bibinfo {author} {\bibfnamefont {M.}~\bibnamefont {Cotallo}},
  \bibinfo {author} {\bibfnamefont {A.}~\bibnamefont {Cruz}}, \bibinfo {author}
  {\bibfnamefont {L.~A.}\ \bibnamefont {Fernandez}}, \bibinfo {author}
  {\bibfnamefont {A.}~\bibnamefont {Gordillo}}, \bibinfo {author}
  {\bibfnamefont {A.}~\bibnamefont {Maiorano}}, \bibinfo {author}
  {\bibfnamefont {F.}~\bibnamefont {Mantovani}}, \bibinfo {author}
  {\bibfnamefont {E.}~\bibnamefont {Marinari}}, \bibinfo {author}
  {\bibfnamefont {V.}~\bibnamefont {Martin-Mayor}}, \bibinfo {author}
  {\bibfnamefont {A.}~\bibnamefont {Mu{\~n}oz~Sudupe}}, \bibinfo {author}
  {\bibfnamefont {D.}~\bibnamefont {Navarro}}, \bibinfo {author} {\bibfnamefont
  {S.}~\bibnamefont {Perez-Gaviro}}, \bibinfo {author} {\bibfnamefont {J.~J.}\
  \bibnamefont {Ruiz-Lorenzo}}, \bibinfo {author} {\bibfnamefont {S.~F.}\
  \bibnamefont {Schifano}}, \bibinfo {author} {\bibfnamefont {D.}~\bibnamefont
  {Sciretti}}, \bibinfo {author} {\bibfnamefont {A.}~\bibnamefont {Tarancon}},
  \bibinfo {author} {\bibfnamefont {R.}~\bibnamefont {Tripiccione}}, \ and\
  \bibinfo {author} {\bibfnamefont {J.~L.}\ \bibnamefont {Velasco}} (\bibinfo
  {collaboration} {Janus Collaboration}),\ }\href {\doibase
  10.1016/j.cpc.2007.09.006} {\bibfield  {journal} {\bibinfo  {journal} {Comp.
  Phys. Comm.}\ }\textbf {\bibinfo {volume} {178}},\ \bibinfo {pages} {208}
  (\bibinfo {year} {2008}{\natexlab{a}})},\ \Eprint
  {http://arxiv.org/abs/arXiv:0704.3573} {arXiv:0704.3573} \BibitemShut
  {NoStop}%
\bibitem [{\citenamefont {Belletti}\ \emph
  {et~al.}(2009{\natexlab{a}})\citenamefont {Belletti}, \citenamefont
  {Guidetti}, \citenamefont {Maiorano}, \citenamefont {Mantovani},
  \citenamefont {Schifano}, \citenamefont {Tripiccione}, \citenamefont
  {Cotallo}, \citenamefont {Perez-Gaviro}, \citenamefont {Sciretti},
  \citenamefont {Velasco}, \citenamefont {Cruz}, \citenamefont {Navarro},
  \citenamefont {Tarancon}, \citenamefont {Fernandez}, \citenamefont
  {Martin-Mayor}, \citenamefont {Mu{\~n}oz-Sudupe}, \citenamefont {Yllanes},
  \citenamefont {Gordillo-Guerrero}, \citenamefont {Ruiz-Lorenzo},
  \citenamefont {Marinari}, \citenamefont {Parisi}, \citenamefont {Rossi},\
  and\ \citenamefont {Zanier}}]{janus:09}%
  \BibitemOpen
  \bibfield  {author} {\bibinfo {author} {\bibfnamefont {F.}~\bibnamefont
  {Belletti}}, \bibinfo {author} {\bibfnamefont {M.}~\bibnamefont {Guidetti}},
  \bibinfo {author} {\bibfnamefont {A.}~\bibnamefont {Maiorano}}, \bibinfo
  {author} {\bibfnamefont {F.}~\bibnamefont {Mantovani}}, \bibinfo {author}
  {\bibfnamefont {S.~F.}\ \bibnamefont {Schifano}}, \bibinfo {author}
  {\bibfnamefont {R.}~\bibnamefont {Tripiccione}}, \bibinfo {author}
  {\bibfnamefont {M.}~\bibnamefont {Cotallo}}, \bibinfo {author} {\bibfnamefont
  {S.}~\bibnamefont {Perez-Gaviro}}, \bibinfo {author} {\bibfnamefont
  {D.}~\bibnamefont {Sciretti}}, \bibinfo {author} {\bibfnamefont {J.~L.}\
  \bibnamefont {Velasco}}, \bibinfo {author} {\bibfnamefont {A.}~\bibnamefont
  {Cruz}}, \bibinfo {author} {\bibfnamefont {D.}~\bibnamefont {Navarro}},
  \bibinfo {author} {\bibfnamefont {A.}~\bibnamefont {Tarancon}}, \bibinfo
  {author} {\bibfnamefont {L.~A.}\ \bibnamefont {Fernandez}}, \bibinfo {author}
  {\bibfnamefont {V.}~\bibnamefont {Martin-Mayor}}, \bibinfo {author}
  {\bibfnamefont {A.}~\bibnamefont {Mu{\~n}oz-Sudupe}}, \bibinfo {author}
  {\bibfnamefont {D.}~\bibnamefont {Yllanes}}, \bibinfo {author} {\bibfnamefont
  {A.}~\bibnamefont {Gordillo-Guerrero}}, \bibinfo {author} {\bibfnamefont
  {J.~J.}\ \bibnamefont {Ruiz-Lorenzo}}, \bibinfo {author} {\bibfnamefont
  {E.}~\bibnamefont {Marinari}}, \bibinfo {author} {\bibfnamefont
  {G.}~\bibnamefont {Parisi}}, \bibinfo {author} {\bibfnamefont
  {M.}~\bibnamefont {Rossi}}, \ and\ \bibinfo {author} {\bibfnamefont
  {G.}~\bibnamefont {Zanier}} (\bibinfo {collaboration} {Janus
  Collaboration}),\ }\href {\doibase 10.1109/MCSE.2009.11} {\bibfield
  {journal} {\bibinfo  {journal} {Computing in Science and Engineering}\
  }\textbf {\bibinfo {volume} {11}},\ \bibinfo {pages} {48} (\bibinfo {year}
  {2009}{\natexlab{a}})}\BibitemShut {NoStop}%
\bibitem [{\citenamefont {Baity-Jesi}\ \emph {et~al.}(2012)\citenamefont
  {Baity-Jesi}, \citenamefont {Ba\~{n}os}, \citenamefont {Cruz}, \citenamefont
  {Fernandez}, \citenamefont {Gil-Narvion}, \citenamefont {Gordillo-Guerrero},
  \citenamefont {Guidetti}, \citenamefont {Iniguez}, \citenamefont {Maiorano},
  \citenamefont {Mantovani}, \citenamefont {Marinari}, \citenamefont
  {Martin-Mayor}, \citenamefont {Monforte-Garcia}, \citenamefont
  {Munoz~Sudupe}, \citenamefont {Navarro}, \citenamefont {Parisi},
  \citenamefont {Pivanti}, \citenamefont {Perez-Gaviro}, \citenamefont
  {Ricci-Tersenghi}, \citenamefont {Ruiz-Lorenzo}, \citenamefont {Schifano},
  \citenamefont {Seoane}, \citenamefont {Tarancon}, \citenamefont {Tellez},
  \citenamefont {Tripiccione},\ and\ \citenamefont {Yllanes}}]{janus:12b}%
  \BibitemOpen
  \bibfield  {author} {\bibinfo {author} {\bibfnamefont {M.}~\bibnamefont
  {Baity-Jesi}}, \bibinfo {author} {\bibfnamefont {R.~A.}\ \bibnamefont
  {Ba\~{n}os}}, \bibinfo {author} {\bibfnamefont {A.}~\bibnamefont {Cruz}},
  \bibinfo {author} {\bibfnamefont {L.~A.}\ \bibnamefont {Fernandez}}, \bibinfo
  {author} {\bibfnamefont {J.~M.}\ \bibnamefont {Gil-Narvion}}, \bibinfo
  {author} {\bibfnamefont {A.}~\bibnamefont {Gordillo-Guerrero}}, \bibinfo
  {author} {\bibfnamefont {M.}~\bibnamefont {Guidetti}}, \bibinfo {author}
  {\bibfnamefont {D.}~\bibnamefont {Iniguez}}, \bibinfo {author} {\bibfnamefont
  {A.}~\bibnamefont {Maiorano}}, \bibinfo {author} {\bibfnamefont
  {F.}~\bibnamefont {Mantovani}}, \bibinfo {author} {\bibfnamefont
  {E.}~\bibnamefont {Marinari}}, \bibinfo {author} {\bibfnamefont
  {V.}~\bibnamefont {Martin-Mayor}}, \bibinfo {author} {\bibfnamefont
  {J.}~\bibnamefont {Monforte-Garcia}}, \bibinfo {author} {\bibfnamefont
  {A.}~\bibnamefont {Munoz~Sudupe}}, \bibinfo {author} {\bibfnamefont
  {D.}~\bibnamefont {Navarro}}, \bibinfo {author} {\bibfnamefont
  {G.}~\bibnamefont {Parisi}}, \bibinfo {author} {\bibfnamefont
  {M.}~\bibnamefont {Pivanti}}, \bibinfo {author} {\bibfnamefont
  {S.}~\bibnamefont {Perez-Gaviro}}, \bibinfo {author} {\bibfnamefont
  {F.}~\bibnamefont {Ricci-Tersenghi}}, \bibinfo {author} {\bibfnamefont
  {J.~J.}\ \bibnamefont {Ruiz-Lorenzo}}, \bibinfo {author} {\bibfnamefont
  {S.~F.}\ \bibnamefont {Schifano}}, \bibinfo {author} {\bibfnamefont
  {B.}~\bibnamefont {Seoane}}, \bibinfo {author} {\bibfnamefont
  {A.}~\bibnamefont {Tarancon}}, \bibinfo {author} {\bibfnamefont
  {P.}~\bibnamefont {Tellez}}, \bibinfo {author} {\bibfnamefont
  {R.}~\bibnamefont {Tripiccione}}, \ and\ \bibinfo {author} {\bibfnamefont
  {D.}~\bibnamefont {Yllanes}},\ }\href {\doibase 10.1140/epjst/e2012-01636-9}
  {\bibfield  {journal} {\bibinfo  {journal} {Eur. Phys. J. Special Topics}\
  }\textbf {\bibinfo {volume} {{210}}},\ \bibinfo {pages} {{33}} (\bibinfo
  {year} {{2012}})},\ \Eprint {http://arxiv.org/abs/arXiv:1204.4134}
  {arXiv:1204.4134} \BibitemShut {NoStop}%
\bibitem [{\citenamefont {Gunnarsson}\ \emph {et~al.}(1991)\citenamefont
  {Gunnarsson}, \citenamefont {Svedlindh}, \citenamefont {Nordblad},
  \citenamefont {Lundgren}, \citenamefont {Aruga},\ and\ \citenamefont
  {Ito}}]{gunnarsson:91}%
  \BibitemOpen
  \bibfield  {author} {\bibinfo {author} {\bibfnamefont {K.}~\bibnamefont
  {Gunnarsson}}, \bibinfo {author} {\bibfnamefont {P.}~\bibnamefont
  {Svedlindh}}, \bibinfo {author} {\bibfnamefont {P.}~\bibnamefont {Nordblad}},
  \bibinfo {author} {\bibfnamefont {L.}~\bibnamefont {Lundgren}}, \bibinfo
  {author} {\bibfnamefont {H.}~\bibnamefont {Aruga}}, \ and\ \bibinfo {author}
  {\bibfnamefont {A.}~\bibnamefont {Ito}},\ }\href {\doibase
  10.1103/PhysRevB.43.8199} {\bibfield  {journal} {\bibinfo  {journal} {Phys.
  Rev. B}\ }\textbf {\bibinfo {volume} {43}},\ \bibinfo {pages} {8199}
  (\bibinfo {year} {1991})}\BibitemShut {NoStop}%
\bibitem [{\citenamefont {Ballesteros}\ \emph {et~al.}(2000)\citenamefont
  {Ballesteros}, \citenamefont {Cruz}, \citenamefont {Fernandez}, \citenamefont
  {Martin-Mayor}, \citenamefont {Pech}, \citenamefont {Ruiz-Lorenzo},
  \citenamefont {Tarancon}, \citenamefont {Tellez}, \citenamefont {Ullod},\
  and\ \citenamefont {Ungil}}]{ballesteros:00}%
  \BibitemOpen
  \bibfield  {author} {\bibinfo {author} {\bibfnamefont {H.~G.}\ \bibnamefont
  {Ballesteros}}, \bibinfo {author} {\bibfnamefont {A.}~\bibnamefont {Cruz}},
  \bibinfo {author} {\bibfnamefont {L.~A.}\ \bibnamefont {Fernandez}}, \bibinfo
  {author} {\bibfnamefont {V.}~\bibnamefont {Martin-Mayor}}, \bibinfo {author}
  {\bibfnamefont {J.}~\bibnamefont {Pech}}, \bibinfo {author} {\bibfnamefont
  {J.~J.}\ \bibnamefont {Ruiz-Lorenzo}}, \bibinfo {author} {\bibfnamefont
  {A.}~\bibnamefont {Tarancon}}, \bibinfo {author} {\bibfnamefont
  {P.}~\bibnamefont {Tellez}}, \bibinfo {author} {\bibfnamefont {C.~L.}\
  \bibnamefont {Ullod}}, \ and\ \bibinfo {author} {\bibfnamefont
  {C.}~\bibnamefont {Ungil}},\ }\href {\doibase 10.1103/PhysRevB.62.14237}
  {\bibfield  {journal} {\bibinfo  {journal} {Phys. Rev. B}\ }\textbf {\bibinfo
  {volume} {62}},\ \bibinfo {pages} {14237} (\bibinfo {year} {2000})},\ \Eprint
  {http://arxiv.org/abs/arXiv:cond-mat/0006211} {arXiv:cond-mat/0006211}
  \BibitemShut {NoStop}%
\bibitem [{\citenamefont {Palassini}\ and\ \citenamefont
  {Caracciolo}(1999)}]{palassini:99}%
  \BibitemOpen
  \bibfield  {author} {\bibinfo {author} {\bibfnamefont {M.}~\bibnamefont
  {Palassini}}\ and\ \bibinfo {author} {\bibfnamefont {S.}~\bibnamefont
  {Caracciolo}},\ }\href {\doibase 10.1103/PhysRevLett.82.5128} {\bibfield
  {journal} {\bibinfo  {journal} {Phys. Rev. Lett.}\ }\textbf {\bibinfo
  {volume} {82}},\ \bibinfo {pages} {5128} (\bibinfo {year} {1999})},\ \Eprint
  {http://arxiv.org/abs/arXiv:cond-mat/9904246} {arXiv:cond-mat/9904246}
  \BibitemShut {NoStop}%
\bibitem [{\citenamefont {Baity-Jesi}\ \emph {et~al.}(2013)\citenamefont
  {Baity-Jesi}, \citenamefont {Ba\~{n}os}, \citenamefont {Cruz}, \citenamefont
  {Fernandez}, \citenamefont {Gil-Narvion}, \citenamefont {Gordillo-Guerrero},
  \citenamefont {Iniguez}, \citenamefont {Maiorano}, \citenamefont {F.},
  \citenamefont {Marinari}, \citenamefont {Martin-Mayor}, \citenamefont
  {Monforte-Garcia}, \citenamefont {Mu{\~n}oz~Sudupe}, \citenamefont {Navarro},
  \citenamefont {Parisi}, \citenamefont {Perez-Gaviro}, \citenamefont
  {Pivanti}, \citenamefont {Ricci-Tersenghi}, \citenamefont {Ruiz-Lorenzo},
  \citenamefont {Schifano}, \citenamefont {Seoane}, \citenamefont {Tarancon},
  \citenamefont {Tripiccione},\ and\ \citenamefont {Yllanes}}]{janus:13b}%
  \BibitemOpen
  \bibfield  {author} {\bibinfo {author} {\bibfnamefont {M.}~\bibnamefont
  {Baity-Jesi}}, \bibinfo {author} {\bibfnamefont {R.~A.}\ \bibnamefont
  {Ba\~{n}os}}, \bibinfo {author} {\bibfnamefont {A.}~\bibnamefont {Cruz}},
  \bibinfo {author} {\bibfnamefont {L.~A.}\ \bibnamefont {Fernandez}}, \bibinfo
  {author} {\bibfnamefont {J.~M.}\ \bibnamefont {Gil-Narvion}}, \bibinfo
  {author} {\bibfnamefont {A.}~\bibnamefont {Gordillo-Guerrero}}, \bibinfo
  {author} {\bibfnamefont {D.}~\bibnamefont {Iniguez}}, \bibinfo {author}
  {\bibfnamefont {A.}~\bibnamefont {Maiorano}}, \bibinfo {author}
  {\bibfnamefont {M.}~\bibnamefont {F.}}, \bibinfo {author} {\bibfnamefont
  {E.}~\bibnamefont {Marinari}}, \bibinfo {author} {\bibfnamefont
  {V.}~\bibnamefont {Martin-Mayor}}, \bibinfo {author} {\bibfnamefont
  {J.}~\bibnamefont {Monforte-Garcia}}, \bibinfo {author} {\bibfnamefont
  {A.}~\bibnamefont {Mu{\~n}oz~Sudupe}}, \bibinfo {author} {\bibfnamefont
  {D.}~\bibnamefont {Navarro}}, \bibinfo {author} {\bibfnamefont
  {G.}~\bibnamefont {Parisi}}, \bibinfo {author} {\bibfnamefont
  {S.}~\bibnamefont {Perez-Gaviro}}, \bibinfo {author} {\bibfnamefont
  {M.}~\bibnamefont {Pivanti}}, \bibinfo {author} {\bibfnamefont
  {F.}~\bibnamefont {Ricci-Tersenghi}}, \bibinfo {author} {\bibfnamefont
  {J.~J.}\ \bibnamefont {Ruiz-Lorenzo}}, \bibinfo {author} {\bibfnamefont
  {S.~F.}\ \bibnamefont {Schifano}}, \bibinfo {author} {\bibfnamefont
  {B.}~\bibnamefont {Seoane}}, \bibinfo {author} {\bibfnamefont
  {A.}~\bibnamefont {Tarancon}}, \bibinfo {author} {\bibfnamefont
  {R.}~\bibnamefont {Tripiccione}}, \ and\ \bibinfo {author} {\bibfnamefont
  {D.}~\bibnamefont {Yllanes}} (\bibinfo {collaboration} {Janus
  Collaboration}),\ }\href@noop {} {\bibfield  {journal} {\bibinfo  {journal}
  {Phys. Rev. B (in press)}\ } (\bibinfo {year} {{2013}})},\ \Eprint
  {http://arxiv.org/abs/arXiv:1310.2910} {arXiv:1310.2910} \BibitemShut
  {NoStop}%
\bibitem [{\citenamefont {Billoire}\ \emph {et~al.}(2012)\citenamefont
  {Billoire}, \citenamefont {Maiorano},\ and\ \citenamefont
  {Marinari}}]{billoire:12}%
  \BibitemOpen
  \bibfield  {author} {\bibinfo {author} {\bibfnamefont {A.}~\bibnamefont
  {Billoire}}, \bibinfo {author} {\bibfnamefont {A.}~\bibnamefont {Maiorano}},
  \ and\ \bibinfo {author} {\bibfnamefont {E.}~\bibnamefont {Marinari}},\
  }\href {\doibase 10.1088/1742-5468/2012/12/P12008} {\bibfield  {journal}
  {\bibinfo  {journal} {J. Stat. Mech.}\ ,\ \bibinfo {pages} {P12008}}
  (\bibinfo {year} {2012})},\ \Eprint {http://arxiv.org/abs/arXiv:1205.2759}
  {arXiv:1205.2759} \BibitemShut {NoStop}%
\bibitem [{\citenamefont {I{\~n}iguez}\ \emph {et~al.}(1996)\citenamefont
  {I{\~n}iguez}, \citenamefont {Parisi},\ and\ \citenamefont
  {Ruiz-Lorenzo}}]{iniguez:96}%
  \BibitemOpen
  \bibfield  {author} {\bibinfo {author} {\bibfnamefont {D.}~\bibnamefont
  {I{\~n}iguez}}, \bibinfo {author} {\bibfnamefont {G.}~\bibnamefont {Parisi}},
  \ and\ \bibinfo {author} {\bibfnamefont {J.~J.}\ \bibnamefont
  {Ruiz-Lorenzo}},\ }\href {\doibase 10.1088/0305-4470/29/15/009} {\bibfield
  {journal} {\bibinfo  {journal} {J. Phys. A: Math. and Gen.}\ }\textbf
  {\bibinfo {volume} {29}},\ \bibinfo {pages} {4337} (\bibinfo {year}
  {1996})},\ \Eprint {http://arxiv.org/abs/cond-mat/9603083} {cond-mat/9603083}
  \BibitemShut {NoStop}%
\bibitem [{\citenamefont {Aizenman}\ and\ \citenamefont
  {Contucci}(1998)}]{aizenman:98}%
  \BibitemOpen
  \bibfield  {author} {\bibinfo {author} {\bibfnamefont {M.}~\bibnamefont
  {Aizenman}}\ and\ \bibinfo {author} {\bibfnamefont {P.}~\bibnamefont
  {Contucci}},\ }\href {\doibase 10.1023/A:1023080223894} {\bibfield  {journal}
  {\bibinfo  {journal} {J. Stat. Phys.}\ }\textbf {\bibinfo {volume} {92}},\
  \bibinfo {pages} {765} (\bibinfo {year} {1998})},\ \Eprint
  {http://arxiv.org/abs/arXiv:cond-mat/9712129} {arXiv:cond-mat/9712129}
  \BibitemShut {NoStop}%
\bibitem [{\citenamefont {Ghirlanda}\ and\ \citenamefont
  {Guerra}(1998)}]{ghirlanda:98}%
  \BibitemOpen
  \bibfield  {author} {\bibinfo {author} {\bibfnamefont {S.}~\bibnamefont
  {Ghirlanda}}\ and\ \bibinfo {author} {\bibfnamefont {F.}~\bibnamefont
  {Guerra}},\ }\href {\doibase 10.1088/0305-4470/31/46/006} {\bibfield
  {journal} {\bibinfo  {journal} {J. Phys. A: Math. Gen.}\ }\textbf {\bibinfo
  {volume} {31}},\ \bibinfo {pages} {9149} (\bibinfo {year} {1998})},\ \Eprint
  {http://arxiv.org/abs/arXiv:cond-mat/9807333} {arXiv:cond-mat/9807333}
  \BibitemShut {NoStop}%
\bibitem [{\citenamefont {Parisi}(1998)}]{parisi:98}%
  \BibitemOpen
  \bibfield  {author} {\bibinfo {author} {\bibfnamefont {G.}~\bibnamefont
  {Parisi}},\ }\href@noop {} {\  (\bibinfo {year} {1998})},\ \Eprint
  {http://arxiv.org/abs/cond-mat/9801081} {cond-mat/9801081} \BibitemShut
  {NoStop}%
\bibitem [{\citenamefont {Contucci}(2003)}]{contucci:03}%
  \BibitemOpen
  \bibfield  {author} {\bibinfo {author} {\bibfnamefont {P.}~\bibnamefont
  {Contucci}},\ }\href {\doibase 10.1088/0305-4470/36/43/020} {\bibfield
  {journal} {\bibinfo  {journal} {J. Phys. A: Math. Gen.}\ }\textbf {\bibinfo
  {volume} {36}},\ \bibinfo {pages} {10961} (\bibinfo {year} {2003})},\ \Eprint
  {http://arxiv.org/abs/arXiv:cond-mat/0302500} {arXiv:cond-mat/0302500}
  \BibitemShut {NoStop}%
\bibitem [{\citenamefont {Ba\~nos}\ \emph {et~al.}(2011)\citenamefont
  {Ba\~nos}, \citenamefont {Cruz}, \citenamefont {Fernandez}, \citenamefont
  {Gil-Narvion}, \citenamefont {Gordillo-Guerrero}, \citenamefont {Guidetti},
  \citenamefont {I\~niguez}, \citenamefont {Maiorano}, \citenamefont
  {Mantovani}, \citenamefont {Marinari}, \citenamefont {Martin-Mayor},
  \citenamefont {Monforte-Garcia}, \citenamefont {Mu\~noz Sudupe},
  \citenamefont {Navarro}, \citenamefont {Parisi}, \citenamefont
  {Perez-Gaviro}, \citenamefont {Ricci-Tersenghi}, \citenamefont
  {Ruiz-Lorenzo}, \citenamefont {Schifano}, \citenamefont {Seoane},
  \citenamefont {Taranc\'on}, \citenamefont {Tripiccione},\ and\ \citenamefont
  {Yllanes}}]{janus:11}%
  \BibitemOpen
  \bibfield  {author} {\bibinfo {author} {\bibfnamefont {R.~A.}\ \bibnamefont
  {Ba\~nos}}, \bibinfo {author} {\bibfnamefont {A.}~\bibnamefont {Cruz}},
  \bibinfo {author} {\bibfnamefont {L.~A.}\ \bibnamefont {Fernandez}}, \bibinfo
  {author} {\bibfnamefont {J.~M.}\ \bibnamefont {Gil-Narvion}}, \bibinfo
  {author} {\bibfnamefont {A.}~\bibnamefont {Gordillo-Guerrero}}, \bibinfo
  {author} {\bibfnamefont {M.}~\bibnamefont {Guidetti}}, \bibinfo {author}
  {\bibfnamefont {D.}~\bibnamefont {I\~niguez}}, \bibinfo {author}
  {\bibfnamefont {A.}~\bibnamefont {Maiorano}}, \bibinfo {author}
  {\bibfnamefont {F.}~\bibnamefont {Mantovani}}, \bibinfo {author}
  {\bibfnamefont {E.}~\bibnamefont {Marinari}}, \bibinfo {author}
  {\bibfnamefont {V.}~\bibnamefont {Martin-Mayor}}, \bibinfo {author}
  {\bibfnamefont {J.}~\bibnamefont {Monforte-Garcia}}, \bibinfo {author}
  {\bibfnamefont {A.}~\bibnamefont {Mu\~noz Sudupe}}, \bibinfo {author}
  {\bibfnamefont {D.}~\bibnamefont {Navarro}}, \bibinfo {author} {\bibfnamefont
  {G.}~\bibnamefont {Parisi}}, \bibinfo {author} {\bibfnamefont
  {S.}~\bibnamefont {Perez-Gaviro}}, \bibinfo {author} {\bibfnamefont
  {F.}~\bibnamefont {Ricci-Tersenghi}}, \bibinfo {author} {\bibfnamefont
  {J.~J.}\ \bibnamefont {Ruiz-Lorenzo}}, \bibinfo {author} {\bibfnamefont
  {S.~F.}\ \bibnamefont {Schifano}}, \bibinfo {author} {\bibfnamefont
  {B.}~\bibnamefont {Seoane}}, \bibinfo {author} {\bibfnamefont
  {A.}~\bibnamefont {Taranc\'on}}, \bibinfo {author} {\bibfnamefont
  {R.}~\bibnamefont {Tripiccione}}, \ and\ \bibinfo {author} {\bibfnamefont
  {D.}~\bibnamefont {Yllanes}},\ }\href {\doibase 10.1103/PhysRevB.84.174209}
  {\bibfield  {journal} {\bibinfo  {journal} {Phys. Rev. B}\ }\textbf {\bibinfo
  {volume} {84}},\ \bibinfo {pages} {174209} (\bibinfo {year} {2011})},\
  \Eprint {http://arxiv.org/abs/arXiv:1107.5772} {arXiv:1107.5772} \BibitemShut
  {NoStop}%
\bibitem [{\citenamefont {Panchenko}(2013)}]{panchenko:13}%
  \BibitemOpen
  \bibfield  {author} {\bibinfo {author} {\bibfnamefont {D.}~\bibnamefont
  {Panchenko}},\ }\href@noop {} {\bibfield  {journal} {\bibinfo  {journal}
  {Ann. of Math.}\ }\textbf {\bibinfo {volume} {177}},\ \bibinfo {pages} {383}
  (\bibinfo {year} {2013})},\ \Eprint {http://arxiv.org/abs/arXiv:1112.1003}
  {arXiv:1112.1003} \BibitemShut {NoStop}%
\bibitem [{\citenamefont {de~Dominicis}\ \emph {et~al.}(1998)\citenamefont
  {de~Dominicis}, \citenamefont {Kondor},\ and\ \citenamefont
  {Temesv{\'a}ri}}]{dedominicis:98}%
  \BibitemOpen
  \bibfield  {author} {\bibinfo {author} {\bibfnamefont {C.}~\bibnamefont
  {de~Dominicis}}, \bibinfo {author} {\bibfnamefont {I.}~\bibnamefont
  {Kondor}}, \ and\ \bibinfo {author} {\bibfnamefont {T.}~\bibnamefont
  {Temesv{\'a}ri}},\ }in\ \href@noop {} {\emph {\bibinfo {booktitle} {{Spin
  {G}lasses and {R}andom {F}ields}}}},\ \bibinfo {editor} {edited by\ \bibinfo
  {editor} {\bibfnamefont {A.~P.}\ \bibnamefont {Young}}}\ (\bibinfo
  {publisher} {World Scientific},\ \bibinfo {address} {Singapore},\ \bibinfo
  {year} {1998})\BibitemShut {NoStop}%
\bibitem [{\citenamefont {Temesv{\'a}ri}\ and\ \citenamefont
  {de~Dominicis}(2002)}]{temesvari:02}%
  \BibitemOpen
  \bibfield  {author} {\bibinfo {author} {\bibfnamefont {T.}~\bibnamefont
  {Temesv{\'a}ri}}\ and\ \bibinfo {author} {\bibfnamefont {C.}~\bibnamefont
  {de~Dominicis}},\ }\href {\doibase 10.1103/PhysRevLett.89.097204} {\bibfield
  {journal} {\bibinfo  {journal} {Phys. Rev. Lett.}\ }\textbf {\bibinfo
  {volume} {89}},\ \bibinfo {pages} {097204} (\bibinfo {year} {2002})},\
  \Eprint {http://arxiv.org/abs/arXiv:cond-mat/0207512}
  {arXiv:cond-mat/0207512} \BibitemShut {NoStop}%
\bibitem [{\citenamefont {de~Dominicis}\ and\ \citenamefont
  {Giardina}(2006)}]{dedominicis:06}%
  \BibitemOpen
  \bibfield  {author} {\bibinfo {author} {\bibfnamefont {C.}~\bibnamefont
  {de~Dominicis}}\ and\ \bibinfo {author} {\bibfnamefont {I.}~\bibnamefont
  {Giardina}},\ }\href@noop {} {\emph {\bibinfo {title} {{Random {F}ields and
  {S}pin {G}lasses}}}}\ (\bibinfo  {publisher} {Cambridge University Press},\
  \bibinfo {address} {Cambridge, England},\ \bibinfo {year} {2006})\BibitemShut
  {NoStop}%
\bibitem [{\citenamefont {de~Dominicis}\ \emph {et~al.}(1993)\citenamefont
  {de~Dominicis}, \citenamefont {Kondor},\ and\ \citenamefont
  {Temesv{\'a}ri}}]{dedominicis:93}%
  \BibitemOpen
  \bibfield  {author} {\bibinfo {author} {\bibfnamefont {C.}~\bibnamefont
  {de~Dominicis}}, \bibinfo {author} {\bibfnamefont {I.}~\bibnamefont
  {Kondor}}, \ and\ \bibinfo {author} {\bibfnamefont {T.}~\bibnamefont
  {Temesv{\'a}ri}},\ }\href {\doibase 10.1142/S0217979293002134} {\bibfield
  {journal} {\bibinfo  {journal} {Int. J. Mod. Phys. B}\ }\textbf {\bibinfo
  {volume} {7}},\ \bibinfo {pages} {986} (\bibinfo {year} {1993})}\BibitemShut
  {NoStop}%
\bibitem [{\citenamefont {Parisi}\ \emph {et~al.}(1997)\citenamefont {Parisi},
  \citenamefont {Ranieri}, \citenamefont {Ricci-Tersenghi},\ and\ \citenamefont
  {Ruiz-Lorenzo}}]{parisi:97}%
  \BibitemOpen
  \bibfield  {author} {\bibinfo {author} {\bibfnamefont {G.}~\bibnamefont
  {Parisi}}, \bibinfo {author} {\bibfnamefont {P.}~\bibnamefont {Ranieri}},
  \bibinfo {author} {\bibfnamefont {F.}~\bibnamefont {Ricci-Tersenghi}}, \ and\
  \bibinfo {author} {\bibfnamefont {J.~J.}\ \bibnamefont {Ruiz-Lorenzo}},\
  }\href {\doibase 10.1088/0305-4470/30/20/015} {\bibfield  {journal} {\bibinfo
   {journal} {J. Phys A: Math. Gen.}\ }\textbf {\bibinfo {volume} {30}},\
  \bibinfo {pages} {7115} (\bibinfo {year} {1997})},\ \Eprint
  {http://arxiv.org/abs/arXiv:cond-mat/9702030} {arXiv:cond-mat/9702030}
  \BibitemShut {NoStop}%
\bibitem [{\citenamefont {Boettcher}(2004)}]{boettcher:04}%
  \BibitemOpen
  \bibfield  {author} {\bibinfo {author} {\bibfnamefont {S.}~\bibnamefont
  {Boettcher}},\ }\href {\doibase 10.1140/epjb/e2004-00102-5} {\bibfield
  {journal} {\bibinfo  {journal} {Eur. Phys. J. B}\ }\textbf {\bibinfo {volume}
  {38}},\ \bibinfo {pages} {83} (\bibinfo {year} {2004})},\ \Eprint
  {http://arxiv.org/abs/arXiv:cond-mat/0310698} {arXiv:cond-mat/0310698}
  \BibitemShut {NoStop}%
\bibitem [{\citenamefont {Yllanes}(2011)}]{yllanes:11}%
  \BibitemOpen
  \bibfield  {author} {\bibinfo {author} {\bibfnamefont {D.}~\bibnamefont
  {Yllanes}},\ }\href@noop {} {\emph {\bibinfo {title} {Rugged Free-Energy
  Landscapes in Disordered Spin Systems}}}\ (\bibinfo  {publisher} {Ph.D.
  thesis, UCM},\ \bibinfo {year} {2011})\ \Eprint
  {http://arxiv.org/abs/arXiv:1111.0266} {arXiv:1111.0266} \BibitemShut
  {NoStop}%
\bibitem [{\citenamefont {Belletti}\ \emph
  {et~al.}(2008{\natexlab{b}})\citenamefont {Belletti}, \citenamefont
  {Cotallo}, \citenamefont {Cruz}, \citenamefont {Fernandez}, \citenamefont
  {Gordillo-Guerrero}, \citenamefont {Guidetti}, \citenamefont {Maiorano},
  \citenamefont {Mantovani}, \citenamefont {Marinari}, \citenamefont
  {Martin-Mayor}, \citenamefont {Mu{\~n}oz~Sudupe}, \citenamefont {Navarro},
  \citenamefont {Parisi}, \citenamefont {Perez-Gaviro}, \citenamefont
  {Ruiz-Lorenzo}, \citenamefont {Schifano}, \citenamefont {Sciretti},
  \citenamefont {Tarancon}, \citenamefont {Tripiccione}, \citenamefont
  {Velasco},\ and\ \citenamefont {Yllanes}}]{janus:08b}%
  \BibitemOpen
  \bibfield  {author} {\bibinfo {author} {\bibfnamefont {F.}~\bibnamefont
  {Belletti}}, \bibinfo {author} {\bibfnamefont {M.}~\bibnamefont {Cotallo}},
  \bibinfo {author} {\bibfnamefont {A.}~\bibnamefont {Cruz}}, \bibinfo {author}
  {\bibfnamefont {L.~A.}\ \bibnamefont {Fernandez}}, \bibinfo {author}
  {\bibfnamefont {A.}~\bibnamefont {Gordillo-Guerrero}}, \bibinfo {author}
  {\bibfnamefont {M.}~\bibnamefont {Guidetti}}, \bibinfo {author}
  {\bibfnamefont {A.}~\bibnamefont {Maiorano}}, \bibinfo {author}
  {\bibfnamefont {F.}~\bibnamefont {Mantovani}}, \bibinfo {author}
  {\bibfnamefont {E.}~\bibnamefont {Marinari}}, \bibinfo {author}
  {\bibfnamefont {V.}~\bibnamefont {Martin-Mayor}}, \bibinfo {author}
  {\bibfnamefont {A.}~\bibnamefont {Mu{\~n}oz~Sudupe}}, \bibinfo {author}
  {\bibfnamefont {D.}~\bibnamefont {Navarro}}, \bibinfo {author} {\bibfnamefont
  {G.}~\bibnamefont {Parisi}}, \bibinfo {author} {\bibfnamefont
  {S.}~\bibnamefont {Perez-Gaviro}}, \bibinfo {author} {\bibfnamefont {J.~J.}\
  \bibnamefont {Ruiz-Lorenzo}}, \bibinfo {author} {\bibfnamefont {S.~F.}\
  \bibnamefont {Schifano}}, \bibinfo {author} {\bibfnamefont {D.}~\bibnamefont
  {Sciretti}}, \bibinfo {author} {\bibfnamefont {A.}~\bibnamefont {Tarancon}},
  \bibinfo {author} {\bibfnamefont {R.}~\bibnamefont {Tripiccione}}, \bibinfo
  {author} {\bibfnamefont {J.~L.}\ \bibnamefont {Velasco}}, \ and\ \bibinfo
  {author} {\bibfnamefont {D.}~\bibnamefont {Yllanes}} (\bibinfo
  {collaboration} {Janus Collaboration}),\ }\href {\doibase
  10.1103/PhysRevLett.101.157201} {\bibfield  {journal} {\bibinfo  {journal}
  {Phys. Rev. Lett.}\ }\textbf {\bibinfo {volume} {101}},\ \bibinfo {pages}
  {157201} (\bibinfo {year} {2008}{\natexlab{b}})},\ \Eprint
  {http://arxiv.org/abs/arXiv:0804.1471} {arXiv:0804.1471} \BibitemShut
  {NoStop}%
\bibitem [{\citenamefont {Belletti}\ \emph
  {et~al.}(2009{\natexlab{b}})\citenamefont {Belletti}, \citenamefont {Cruz},
  \citenamefont {Fernandez}, \citenamefont {Gordillo-Guerrero}, \citenamefont
  {Guidetti}, \citenamefont {Maiorano}, \citenamefont {Mantovani},
  \citenamefont {Marinari}, \citenamefont {Martin-Mayor}, \citenamefont
  {Monforte}, \citenamefont {Mu{\~n}oz~Sudupe}, \citenamefont {Navarro},
  \citenamefont {Parisi}, \citenamefont {Perez-Gaviro}, \citenamefont
  {Ruiz-Lorenzo}, \citenamefont {Schifano}, \citenamefont {Sciretti},
  \citenamefont {Tarancon}, \citenamefont {Tripiccione},\ and\ \citenamefont
  {Yllanes}}]{janus:09b}%
  \BibitemOpen
  \bibfield  {author} {\bibinfo {author} {\bibfnamefont {F.}~\bibnamefont
  {Belletti}}, \bibinfo {author} {\bibfnamefont {A.}~\bibnamefont {Cruz}},
  \bibinfo {author} {\bibfnamefont {L.~A.}\ \bibnamefont {Fernandez}}, \bibinfo
  {author} {\bibfnamefont {A.}~\bibnamefont {Gordillo-Guerrero}}, \bibinfo
  {author} {\bibfnamefont {M.}~\bibnamefont {Guidetti}}, \bibinfo {author}
  {\bibfnamefont {A.}~\bibnamefont {Maiorano}}, \bibinfo {author}
  {\bibfnamefont {F.}~\bibnamefont {Mantovani}}, \bibinfo {author}
  {\bibfnamefont {E.}~\bibnamefont {Marinari}}, \bibinfo {author}
  {\bibfnamefont {V.}~\bibnamefont {Martin-Mayor}}, \bibinfo {author}
  {\bibfnamefont {J.}~\bibnamefont {Monforte}}, \bibinfo {author}
  {\bibfnamefont {A.}~\bibnamefont {Mu{\~n}oz~Sudupe}}, \bibinfo {author}
  {\bibfnamefont {D.}~\bibnamefont {Navarro}}, \bibinfo {author} {\bibfnamefont
  {G.}~\bibnamefont {Parisi}}, \bibinfo {author} {\bibfnamefont
  {S.}~\bibnamefont {Perez-Gaviro}}, \bibinfo {author} {\bibfnamefont {J.~J.}\
  \bibnamefont {Ruiz-Lorenzo}}, \bibinfo {author} {\bibfnamefont {S.~F.}\
  \bibnamefont {Schifano}}, \bibinfo {author} {\bibfnamefont {D.}~\bibnamefont
  {Sciretti}}, \bibinfo {author} {\bibfnamefont {A.}~\bibnamefont {Tarancon}},
  \bibinfo {author} {\bibfnamefont {R.}~\bibnamefont {Tripiccione}}, \ and\
  \bibinfo {author} {\bibfnamefont {D.}~\bibnamefont {Yllanes}} (\bibinfo
  {collaboration} {Janus Collaboration}),\ }\href {\doibase
  10.1007/s10955-009-9727-z} {\bibfield  {journal} {\bibinfo  {journal} {J.
  Stat. Phys.}\ }\textbf {\bibinfo {volume} {135}},\ \bibinfo {pages} {1121}
  (\bibinfo {year} {2009}{\natexlab{b}})},\ \Eprint
  {http://arxiv.org/abs/arXiv:0811.2864} {arXiv:0811.2864} \BibitemShut
  {NoStop}%
\end{thebibliography}
\end{document}